\documentclass[twocolumn,twoside,slac_two]{revtex4}
\usepackage{graphicx}
\usepackage{fancyhdr}
\usepackage{subfigure}
\usepackage{epstopdf}
\begin{document}

\title{An alternative solution to the $\gamma$-ray Gradient problem}

\author{D. Gaggero}
\affiliation{INFN Pisa and Pisa University, Largo B. Pontecorvo 3, I-56127 Pisa, Italy}

\author{C. Evoli}
\affiliation{Institut f\"{u}r Theoretische Physik, Universit\"{a}t Hamburg, Luruper Chaussee 149, D-22761 Hamburg, Germany.}

\author{D. Grasso}
\affiliation{INFN Pisa, Largo B. Pontecorvo 3, I-56127 Pisa, Italy}

\author{L. Maccione}
\affiliation{Ludwig-Maximilians-Universit\"{a}t, Fakult\"{a}t f\"{u}r Physik, Theresienstra\ss e 37, D-80333 M\"{u}nchen, Germany}
\affiliation{Max-Planck-Institut f\"{u}r Physik, Werner-Heisenberg-Institut, F\"{o}hringer Ring 6, D-80805 M\"{u}nchen, Germany}

\begin{abstract}
The Fermi-LAT collaboration recently confirmed EGRET finding of a discrepancy between the observed longitudinal profile
of $\gamma$-ray diffuse emission from the Galaxy and that computed with {\tt GALPROP} assuming that cosmic rays are produced by
Galactic supernova remnants. The accurate Fermi-LAT measurements make this anomaly hardly explainable in terms of conventional diffusion schemes.
Here we use {\tt DRAGON} numerical diffusion code to implement a physically motivated scenario in which the diffusion coefficient is spatially correlated to the source density.
We show that under those conditions we are able to reproduce the observed flat emissivity profile in the outer Galaxy with no need to change the source term, 
the  diffusion halo height, or the CO-${\rm H_2}$ conversion factor (${\rm X_{CO}}$) with respect to  
their preferred values/distributions. We also show that our models are compatible with gamma-ray longitudinal   
profiles measured by Fermi-LAT, and still provide a satisfactory fit of all observed secondary-to-primary ratios, such as B/C and antiprotons/protons. 
\end{abstract}

\maketitle

\thispagestyle{fancy}


\section{Introduction}

It has been known since the EGRET era that, if one computes the cosmic ray (CR) Galactocentric radial distribution adopting a source function deduced from pulsar or supernova remnant (SNR) catalogues, the result appears much steeper than the profile inferred from the $\gamma$-ray diffuse emission along the Galactic plane: the latter appears flatter, with a high contribution from large Galactic radii. This discrepancy is known as the {\it $\gamma$-ray gradient problem}. A sharp rise of the conversion factor between CO emissivity and ${\rm H_2}$ density (the so called ${\rm X_{CO}}$) with the Galactocentric radius was invoked at the time to fix the problem \citep{gradient_problem_2004}: a larger gas density at large radii compensates for the decreasing CR population and is able to explain the $\gamma$-ray flux detected at high Galactic longitudes.
 
Fermi-LAT confirmed the existence of such a problem \citep{ThirdQuadrant}. Moreover, the high spatial resolution of the LAT permitted to disentangle the emission coming from the interaction of CRs with the molecular gas (whose modelling is strongly affected by the uncertainty on the ${\rm X_{CO}}$) from the emission originated by the interaction of the Galactic CRs with the atomic gas (whose density is better known from its 21 cm radio emission). An analysis based on $\gamma$-ray maps of the third Galactic quadrant \citep{ThirdQuadrant} pointed out that the $\gamma$-ray emissivity from {\it neutral} gas (tracing the actual CR density) is indeed flatter than the predicted one confirming the gradient problem independently of the  ${\rm X_{CO}}$. This result led the authors of \citep{ThirdQuadrant} to look for alternative explanations of the problem, e.g. invoking a thick CR diffusion halo or a source term that becomes flatter at large radii.  Both solutions, however, do not appear completely satisfactory: a thick halo is disfavoured both from $^{10}$Be/$^9$Be and synchrotron data; a smooth source distribution is in contrast with SNR catalogues. 

Here we consider a different interpretation based on relaxing the approximation of isotropic and spatially uniform diffusion.

\section{Inhomogeneous and anisotropic diffusion}

Nearly all CR diffusion models presented in the literature adopt an isotropic and spatially uniform diffusion coefficient throughout all the Galaxy.
This is the case, for example, of {\tt GALPROP} numerical package on which the predictions of \citep{ThirdQuadrant} are based. 
It is reasonable, however, to expect that CR diffusion is not isotropic in the Galaxy. 
This could be the consequence either of Galactic winds \citep{Gebauer:2009hk} or just of the anisotropy of the regular component of the Galactic magnetic field which is oriented almost azimutally along the Galactic plane. 
The former possibility was suggested as a possible solution of the $\gamma$-ray gradient problem originated by EGRET observations \citep{Breitschwerdt}.
Here we consider the latter option and extend also to the recent Fermi-LAT data the arguments we developed in [Evoli et al. 2008] to interpret earlier EGRET measurements.
Our approach is based on the consideration that, for geometrical reasons, CRs should escape from the Galaxy almost perpendicularly to the Galactic plane: their density, therefore, should be determined by the perpendicular component of the diffusion coefficient $D_\perp$. 
We know -- both from quasi linear diffusion theory and from more realistic numerical simulations \citep{BlasiDeMarco} -- that $D_\perp$ should {\it increase} with increasing strength of the Galactic magnetic field turbulent component; 
from a physical point of view, such behaviour can be understood in terms of magnetic field line random walk becoming stronger when the turbulence strength increases. 
As a consequence, the regions where CR injection is more intense should also be those characterized by a stronger MHD turbulence and hence  a faster CR escape along the $z$ axis: this should smooth the CR gradient, and hence the $\gamma$-ray profile, in a rather natural way.

In the next section we will show this effect by means of dedicated numerical computations. 


\section{Our method and results}

\begin{figure}[h]
	\setlength{\unitlength}{1mm}
     \begin{center}
     \subfigure{
      \includegraphics[width=5.5cm]{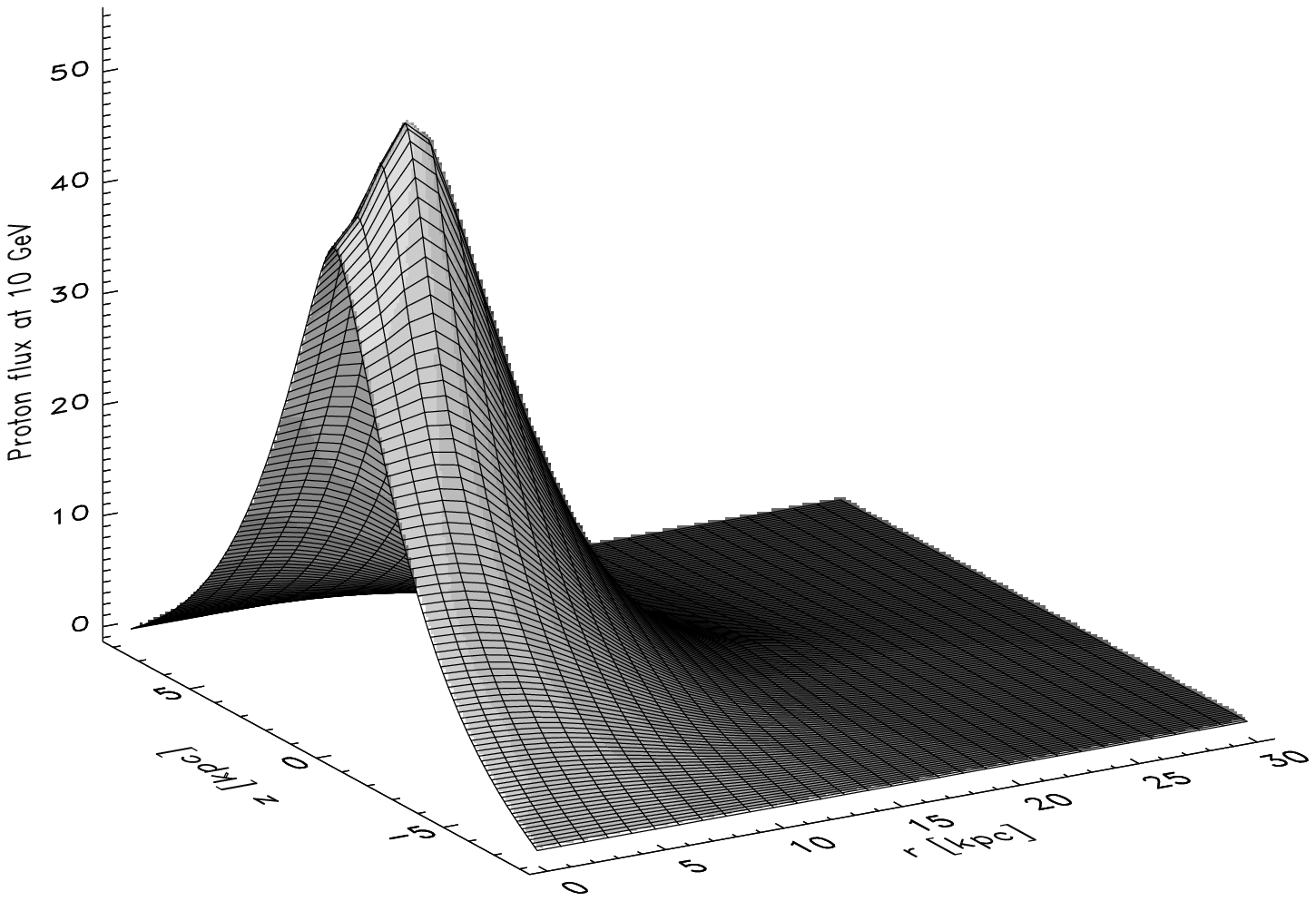}
 	 }
     \subfigure{
      \includegraphics[width=5.5cm]{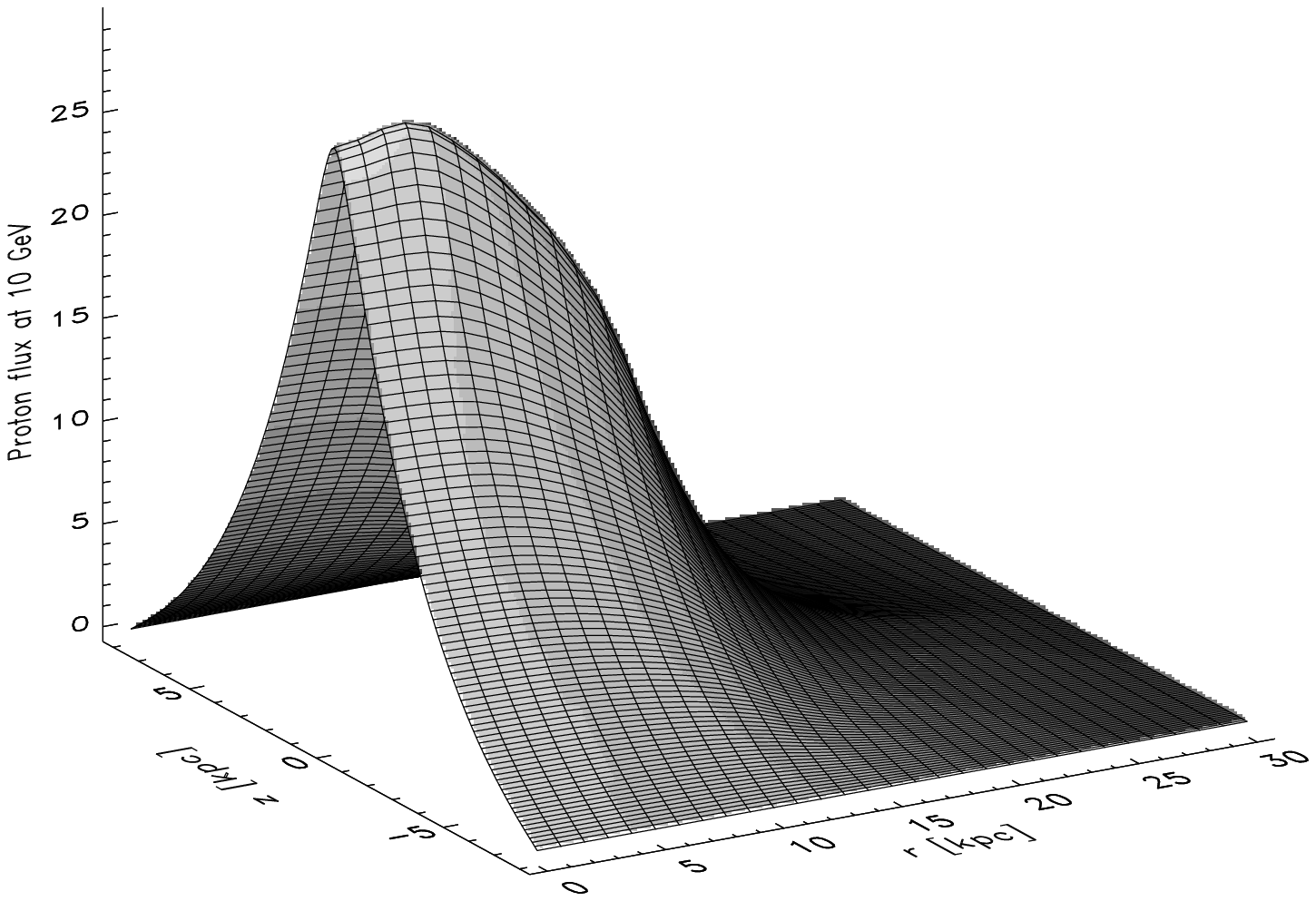}
 	 }
 	 \end{center}
     \caption{\footnotesize \it Two different CR proton distribution maps in arbitrary units computed with {\tt DRAGON} at $10$ GeV are shown as functions of the Galactic cylindrical coordinates $R$ and $z$. {\bf Panel a)} The proton distribution is computed with no radial dependence of diffusion coefficient. {\bf Panel b)} Here the diffusion coefficient is correlated to the source term: $D \propto Q^{\tau}$, with $\tau = 0.8$. The model shows a significant flattening in the CR profile along $R$. The normalization is fixed at $R_{\rm Sun} = 8.5$ kpc in both cases; notice how the maximum proton density is reduced by a factor $\simeq 2$ in the second panel.}
     \label{fig:protons_map}
\end{figure}

\begin{figure}[h]
	\setlength{\unitlength}{1mm}
     \begin{center}
     \subfigure{
      \includegraphics[width=5.5cm]{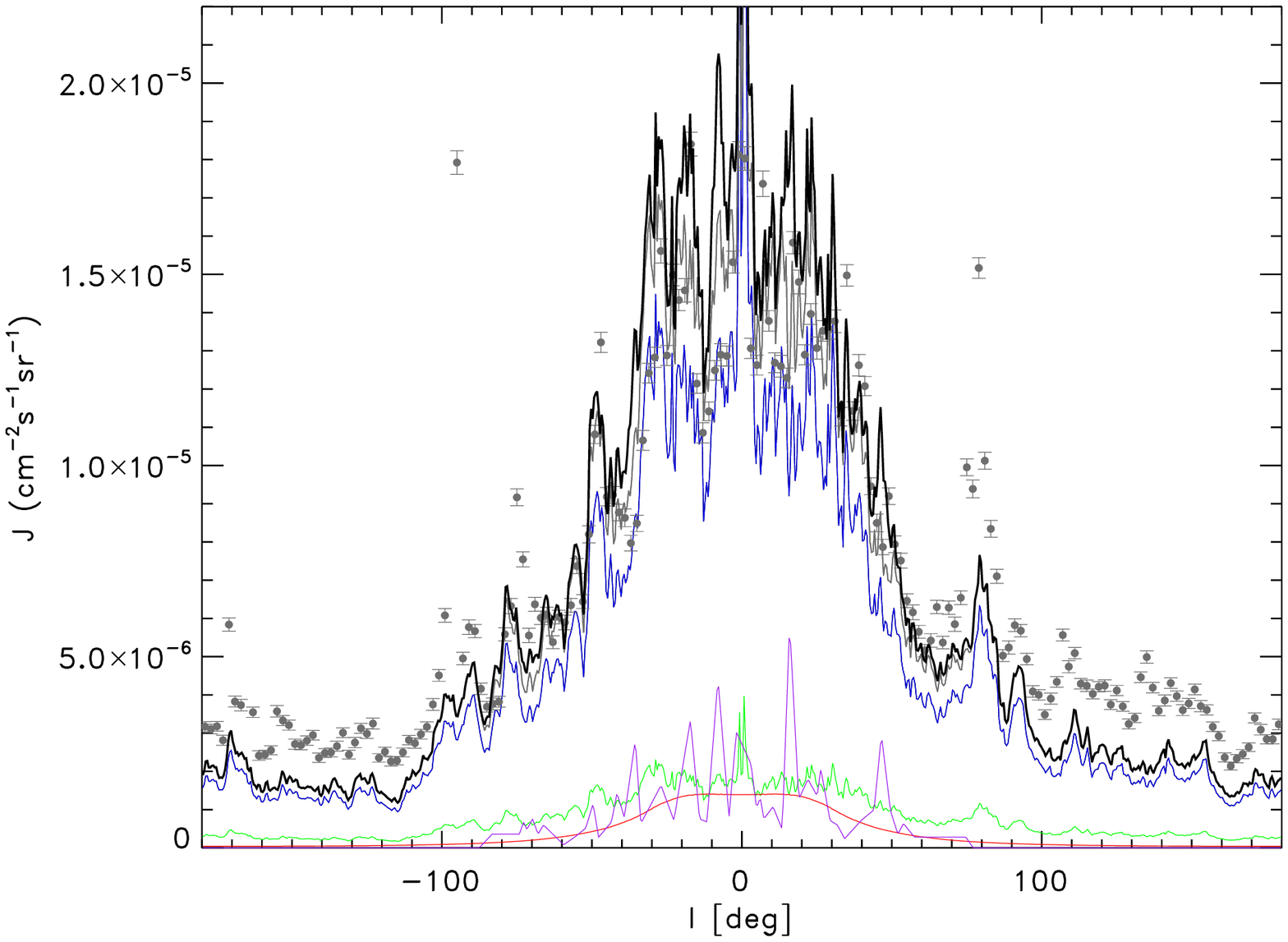}
 	 }
     \subfigure{
      \includegraphics[width=5.5cm]{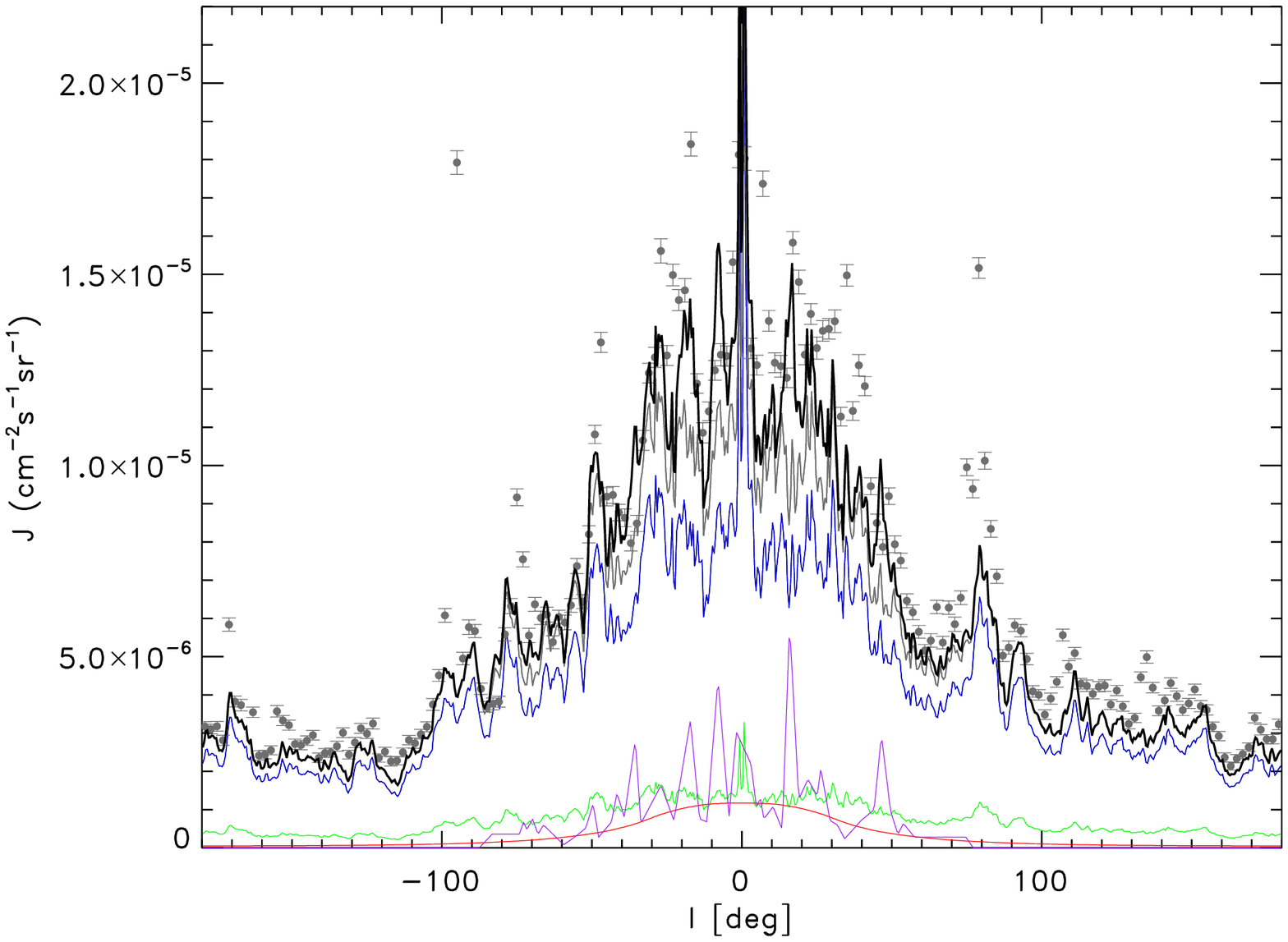}
 	 }
 	 \end{center}
     \caption{\footnotesize \it Two $gamma$-ray longitudinal profiles along the Galactic plane computed with {\tt DRAGON} and {\tt GammaSky} and compared to preliminary Fermi-LAT data extracted from the talk by A.W.Strong at the Workshop on Indirect Dark Matter Searches, DESY, Hamburg, June 2011 ({\tt http://www.mpe.mpg.de/~aws/talks/}). Data are integrated over the latitude interval $-5^{\circ} < b < +5^{\circ}$ and in energy between 1104 and 1442 MeV. Red line: IC. Green line: Bremsstrahlung. Blue line: $\pi^0$ decay. Purple line: contribution from unresolved sources. Grey line: $\pi^0$ + IC + Bremsstrahlung. Black line: total. {\bf Panel a)} The profile is computed with no radial dependence of diffusion coefficient. 
     {\bf Panel b)} Here the diffusion coefficient follows the source term: $D \propto Q^{\tau}$, with $\tau = 0.8$. }
     \label{fig:PD_D_exp_exprad}
\end{figure}

\begin{figure}[t]
\begin{center}
 \centering
 \includegraphics[width = 6. cm]{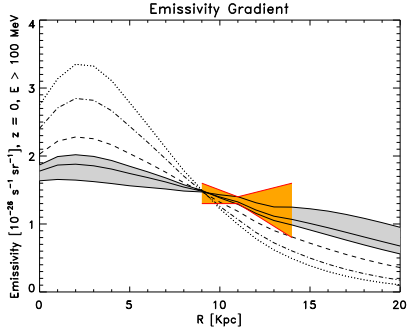}
\caption{\footnotesize \it Here the effect of the parameter $\tau$ defined by equation \ref{D_R} is explored.
Dotted line: no radial dependence of diffusion coefficient ($\tau=0$). Dot-dashed line: $\tau=0.2$. Dashed line: $\tau=0.5$.
Solid lines: $\tau=0.7$ -- $0.8$ -- $1.0$. The values corresponding to the solid lines within the grey band match the observed gradient.}
 \label{fig:PD_exprad_00_02_05_07_08_10_Rmax30}
\end{center}
\end{figure}

In this section we use {\tt DRAGON} numerical diffusion package\footnote{The DRAGON code for cosmic-ray transport and diffuse emission production is available online at \url{http://www.desy.de/~maccione/DRAGON/}} to solve the diffusion equation in the presence of a diffusion coefficient spatially correlated to the CR source term $Q(r,z)$. 
We perform our analysis using a Plain Diffusion (PD) setup with no convection and no reacceleration in order to better highlight the effects of inhomogeneous diffusion; similar results may be obtained with different choices of the diffusion parameters, and a more detailed study on the effects of another setup will be performed in a forthcoming paper.
The CR propagation model adopted here is basically the same as the PD model described in \citep{electrons_final}; the astrophysical parameters (in particular the source term, gas distribution and ${\rm X_{CO}}$) are also the same used in that analysis. Only the normalization 
of the proton injection spectrum is slightly tuned to match 
the recently released proton spectrum measured by the PAMELA collaboration \citep{Pamela_p_2011}.
The model is also compatible with most other CR data sets. 
Only a little excess in the antiproton spectrum must be pointed out which, however, is still compatible with data if astrophysical and particle physics uncertainties are taken into account. 
As we mentioned, the CR distribution is computed with {\tt DRAGON}: this numerical package is suitable for our purpose since, differently from {\tt GALPROP}, it implements the possibility to vary the diffusion coefficient through the Galaxy. The CR distributions is then used as an input to compute the $\gamma$-ray longitude profile along the Galactic plane; the $\gamma$-ray map is evaluated with a separate package called {\tt GammaSky}. 
	
	

The result of a combined {\tt DRAGON} and {\tt GammaSky} computation in the case of a uniform diffusion coefficient and a PD setup is shown in Fig. \ref{fig:PD_D_exp_exprad} (panel a). It is clear from that plot that the predicted longitude profile is too steep compared to the observations: in the Galactic center region the model prediction overshoots the data and in the anti-center region the model is lower than the observations by several $\sigma$.
Tuning the ${\rm X_{CO}}(R)$ could help in principle: assuming a lower value of this parameter in the bulge and a high value at large $R$ could smooth the $\gamma$-ray profile (as done in several previous works such as \citep{gradient_problem_2004}). Unfortunately, as pointed out in the introduction, the gradient problem is present especially in the {\it emissivity profile}, and this quantity is independent of the molecular gas: it only traces the actual CR distribution\footnote{The emissivity is the number of $\gamma$ photons emitted by each gas atom per unit time and unit energy}.
	 	
So we apply our previous considerations and adopt a diffusion coefficient correlated to the radial dependence of the source term $Q(R)$ by the following expression:
	\begin{equation}
	D(R) \, \propto \, Q(R)^{\tau}
	\label{D_R}
	\end{equation}
This is the parametrization we already used in \citep{anstat_2008} to interpret EGRET data. 
The parameter $\tau$ is tuned against data. 
In Fig. \ref{fig:PD_exprad_00_02_05_07_08_10_Rmax30} we show the emissivity profile for different values of $\tau$ in the range ${\rm [0 \div 1]}$. It is evident from that figure that an increasing value of $\tau$ yields a much smoother behaviour of the emissivity as function of $R$. Values in the range ${\rm [0.7 \div 1]}$ allow a good match of Fermi-LAT data (\citep{ThirdQuadrant}, \citep{CasCep}).
	
\begin{figure}[h]
     \setlength{\unitlength}{1mm}
      \begin{center}
     \includegraphics[width = 6. cm]{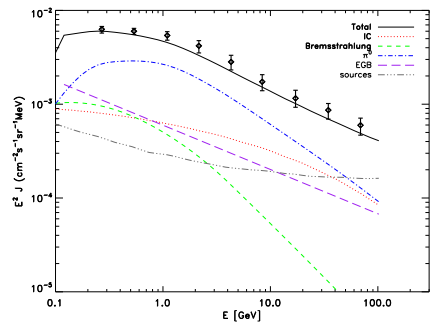}
     \end{center}
     \caption{\footnotesize \it The gamma-ray spectrum corresponding to the plain diffusion model with varying diffusion coefficient described in the text ($\tau = 0.8$). The spectrum was computed with {\tt DRAGON} and {\tt GammaSky}. The data points measured by Fermi-LAT are taken from the same reference as Fig. \ref{fig:PD_D_exp_exprad}}
	\label{fig:spectrum}
\end{figure}

With this result at hand, we considered a modified version of the Plain Diffusion CR propagation setup with $D(R) = Q^{\tau}$ and $\tau = 0.8$. The smoothing in the CR distribution corresponding to such a value of $\tau$ is shown in Fig. \ref{fig:protons_map}. 
As shown in Fig. \ref{fig:PD_D_exp_exprad} (panel b), the $\gamma$ ray longitude profile along the Galactic plane is nicely reproduced {\it with no tuning at all of the ${\rm X_{CO}}$}. It is remarkable that a simple CR propagation setup, with only the addition of the radial dependence of $D$ and no {\it ad hoc} tuning, permits to reproduce the $\gamma$-ray profile with such accuracy.
Noticeably, the modified model is {\it still compatible} with most relevant CR data set, most importantly the B/C. 
Furthermore, we checked that the $\gamma$-ray spectrum measured by Fermi-LAT along the Galactic plane is also correctly reproduced under those conditions (see Fig. \ref{fig:spectrum}).

	
\section{Conclusions}

In this paper we presented an alternative solution to the well known $\gamma$-ray {\it gradient problem}. 
Our approach is based on the physically motivated hypothesis that the CR diffusion coefficient is spatially correlated to the source density: regions in which star, hence SNR, formation is stronger are expected to present a stronger turbulence level and therefore a larger value of the perpendicular diffusion coefficient. This effect favours CR escape from most active regions helping to smooth their density through the Galaxy hence also the $\gamma$-ray gradient.   
We used {\tt DRAGON} package to implement this scenario and to check that CR data are still reproduced under those conditions. 
In spite of being purely phenomenological (as a self-consistent theory/computation of non-linear CR - MHD turbulence interaction in the Galaxy is far from being developed) our approach provides a remarkably good description of the spectrum and longitude distribution of the diffuse $\gamma$-ray emission measured by the Fermi-LAT collaboration.


\begin{acknowledgments}
D. Gaggero would like to thank the LAPTH (Laboratoire d'Annecy-le-Vieux de Physique Th\'{e}orique) for hosting him during the last part of the work presented in this paper.
\end{acknowledgments}

\bigskip 


\bibliographystyle{apsrev}
\bibliography{Proceedings_LaTeX}{}

\begin{thebibliography}{9}
\expandafter\ifx\csname natexlab\endcsname\relax\def\natexlab#1{#1}\fi
\expandafter\ifx\csname bibnamefont\endcsname\relax
  \def\bibnamefont#1{#1}\fi
\expandafter\ifx\csname bibfnamefont\endcsname\relax
  \def\bibfnamefont#1{#1}\fi
\expandafter\ifx\csname citenamefont\endcsname\relax
  \def\citenamefont#1{#1}\fi
\expandafter\ifx\csname url\endcsname\relax
  \def\url#1{\texttt{#1}}\fi
\expandafter\ifx\csname urlprefix\endcsname\relax\def\urlprefix{URL }\fi
\providecommand{\bibinfo}[2]{#2}
\providecommand{\eprint}[2][]{\url{#2}}

\bibitem[{\citenamefont{{Strong} et~al.}(2004)\citenamefont{{Strong},
  {Moskalenko}, {Reimer}, {Digel}, and {Diehl}}}]{gradient_problem_2004}
\bibinfo{author}{\bibfnamefont{A.~W.} \bibnamefont{{Strong}}},
  \bibinfo{author}{\bibfnamefont{I.~V.} \bibnamefont{{Moskalenko}}},
  \bibinfo{author}{\bibfnamefont{O.}~\bibnamefont{{Reimer}}},
  \bibinfo{author}{\bibfnamefont{S.}~\bibnamefont{{Digel}}}, \bibnamefont{and}
  \bibinfo{author}{\bibfnamefont{R.}~\bibnamefont{{Diehl}}},
  \bibinfo{journal}{Astronomy and Astrophysics} \textbf{\bibinfo{volume}{422}},
  \bibinfo{pages}{L47} (\bibinfo{year}{2004}),
  \eprint{\url{http://arxiv.org/abs/astro-ph/0405275}}.

\bibitem[{\citenamefont{{Ackermann} et~al.}(2011)\citenamefont{{Ackermann},
  {Ajello}, {Baldini}, {Ballet}, {Barbiellini}, {Bastieri}, {Bechtol},
  {Bellazzini}, {Berenji}, {Bloom} et~al.}}]{ThirdQuadrant}
\bibinfo{author}{\bibfnamefont{M.}~\bibnamefont{{Ackermann}}},
  \bibinfo{author}{\bibfnamefont{M.}~\bibnamefont{{Ajello}}},
  \bibinfo{author}{\bibfnamefont{L.}~\bibnamefont{{Baldini}}},
  \bibinfo{author}{\bibfnamefont{J.}~\bibnamefont{{Ballet}}},
  \bibinfo{author}{\bibfnamefont{G.}~\bibnamefont{{Barbiellini}}},
  \bibinfo{author}{\bibfnamefont{D.}~\bibnamefont{{Bastieri}}},
  \bibinfo{author}{\bibfnamefont{K.}~\bibnamefont{{Bechtol}}},
  \bibinfo{author}{\bibfnamefont{R.}~\bibnamefont{{Bellazzini}}},
  \bibinfo{author}{\bibfnamefont{B.}~\bibnamefont{{Berenji}}},
  \bibinfo{author}{\bibfnamefont{E.~D.} \bibnamefont{{Bloom}}},
  \bibnamefont{et~al.}, \bibinfo{journal}{Astrophysical Journal}
  \textbf{\bibinfo{volume}{726}}, \bibinfo{pages}{81} (\bibinfo{year}{2011}),
  \eprint{\url{http://arxiv.org/abs/1011.0816}}.

\bibitem[{\citenamefont{Gebauer and de~Boer}(2009)}]{Gebauer:2009hk}
\bibinfo{author}{\bibfnamefont{I.}~\bibnamefont{Gebauer}} \bibnamefont{and}
  \bibinfo{author}{\bibfnamefont{W.}~\bibnamefont{de~Boer}}
  (\bibinfo{year}{2009}), \bibinfo{note}{* Brief entry *},
  \eprint{\url{http://arxiv.org/abs/0910.2027}}.

\bibitem[{\citenamefont{{Breitschwerdt}
  et~al.}(2002)\citenamefont{{Breitschwerdt}, {Dogiel}, and
  {V{\"o}lk}}}]{Breitschwerdt}
\bibinfo{author}{\bibfnamefont{D.}~\bibnamefont{{Breitschwerdt}}},
  \bibinfo{author}{\bibfnamefont{V.~A.} \bibnamefont{{Dogiel}}},
  \bibnamefont{and} \bibinfo{author}{\bibfnamefont{H.~J.}
  \bibnamefont{{V{\"o}lk}}}, \bibinfo{journal}{Astronomy and Astrophysics}
  \textbf{\bibinfo{volume}{385}}, \bibinfo{pages}{216} (\bibinfo{year}{2002}),
  \eprint{\url{http://arxiv.org/abs/astro-ph/0201345}}.

\bibitem[{\citenamefont{{DeMarco} et~al.}(2007)\citenamefont{{DeMarco},
  {Blasi}, and {Stanev}}}]{BlasiDeMarco}
\bibinfo{author}{\bibfnamefont{D.}~\bibnamefont{{DeMarco}}},
  \bibinfo{author}{\bibfnamefont{P.}~\bibnamefont{{Blasi}}}, \bibnamefont{and}
  \bibinfo{author}{\bibfnamefont{T.}~\bibnamefont{{Stanev}}},
  \bibinfo{journal}{Journal of Cosmology and Astroparticle Physics}
  \textbf{\bibinfo{volume}{6}}, \bibinfo{pages}{27} (\bibinfo{year}{2007}),
  \eprint{\url{http://arxiv.org/abs/0705.1972}}.

\bibitem[{\citenamefont{{Di Bernardo} et~al.}(2011)\citenamefont{{Di Bernardo},
  {Evoli}, {Gaggero}, {Grasso}, {Maccione}, and {Mazziotta}}}]{electrons_final}
\bibinfo{author}{\bibfnamefont{G.}~\bibnamefont{{Di Bernardo}}},
  \bibinfo{author}{\bibfnamefont{C.}~\bibnamefont{{Evoli}}},
  \bibinfo{author}{\bibfnamefont{D.}~\bibnamefont{{Gaggero}}},
  \bibinfo{author}{\bibfnamefont{D.}~\bibnamefont{{Grasso}}},
  \bibinfo{author}{\bibfnamefont{L.}~\bibnamefont{{Maccione}}},
  \bibnamefont{and} \bibinfo{author}{\bibfnamefont{M.~N.}
  \bibnamefont{{Mazziotta}}}, \bibinfo{journal}{Astroparticle Physics}
  \textbf{\bibinfo{volume}{34}}, \bibinfo{pages}{528} (\bibinfo{year}{2011}),
  \eprint{\url{http://arxiv.org/pdf/1010.0174v2}}.

\bibitem[{\citenamefont{{Adriani} et~al.}(2011)\citenamefont{{Adriani},
  {Barbarino}, {Bazilevskaya}, {Bellotti}, {Boezio}, {Bogomolov}, {Bonechi},
  {Bongi}, {Bonvicini}, {Borisov} et~al.}}]{Pamela_p_2011}
\bibinfo{author}{\bibfnamefont{O.}~\bibnamefont{{Adriani}}},
  \bibinfo{author}{\bibfnamefont{G.~C.} \bibnamefont{{Barbarino}}},
  \bibinfo{author}{\bibfnamefont{G.~A.} \bibnamefont{{Bazilevskaya}}},
  \bibinfo{author}{\bibfnamefont{R.}~\bibnamefont{{Bellotti}}},
  \bibinfo{author}{\bibfnamefont{M.}~\bibnamefont{{Boezio}}},
  \bibinfo{author}{\bibfnamefont{E.~A.} \bibnamefont{{Bogomolov}}},
  \bibinfo{author}{\bibfnamefont{L.}~\bibnamefont{{Bonechi}}},
  \bibinfo{author}{\bibfnamefont{M.}~\bibnamefont{{Bongi}}},
  \bibinfo{author}{\bibfnamefont{V.}~\bibnamefont{{Bonvicini}}},
  \bibinfo{author}{\bibfnamefont{S.}~\bibnamefont{{Borisov}}},
  \bibnamefont{et~al.}, \bibinfo{journal}{Science}
  \textbf{\bibinfo{volume}{332}}, \bibinfo{pages}{69} (\bibinfo{year}{2011}),
  \eprint{\url{http://arxiv.org/abs/1103.4055}}.

\bibitem[{\citenamefont{{Evoli} et~al.}(2008)\citenamefont{{Evoli}, {Gaggero},
  {Grasso}, and {Maccione}}}]{anstat_2008}
\bibinfo{author}{\bibfnamefont{C.}~\bibnamefont{{Evoli}}},
  \bibinfo{author}{\bibfnamefont{D.}~\bibnamefont{{Gaggero}}},
  \bibinfo{author}{\bibfnamefont{D.}~\bibnamefont{{Grasso}}}, \bibnamefont{and}
  \bibinfo{author}{\bibfnamefont{L.}~\bibnamefont{{Maccione}}},
  \bibinfo{journal}{Journal of Cosmology and Astroparticle Physics}
  \textbf{\bibinfo{volume}{10}}, \bibinfo{pages}{18} (\bibinfo{year}{2008}),
  \eprint{\url{http://arxiv.org/abs/0807.4730}}.

\bibitem[{\citenamefont{{Abdo} et~al.}(2010)\citenamefont{{Abdo}, {Ackermann},
  {Ajello}, {Baldini}, {Ballet}, {Barbiellini}, {Bastieri}, {Baughman},
  {Bechtol}, {Bellazzini} et~al.}}]{CasCep}
\bibinfo{author}{\bibfnamefont{A.~A.} \bibnamefont{{Abdo}}},
  \bibinfo{author}{\bibfnamefont{M.}~\bibnamefont{{Ackermann}}},
  \bibinfo{author}{\bibfnamefont{M.}~\bibnamefont{{Ajello}}},
  \bibinfo{author}{\bibfnamefont{L.}~\bibnamefont{{Baldini}}},
  \bibinfo{author}{\bibfnamefont{J.}~\bibnamefont{{Ballet}}},
  \bibinfo{author}{\bibfnamefont{G.}~\bibnamefont{{Barbiellini}}},
  \bibinfo{author}{\bibfnamefont{D.}~\bibnamefont{{Bastieri}}},
  \bibinfo{author}{\bibfnamefont{B.~M.} \bibnamefont{{Baughman}}},
  \bibinfo{author}{\bibfnamefont{K.}~\bibnamefont{{Bechtol}}},
  \bibinfo{author}{\bibfnamefont{R.}~\bibnamefont{{Bellazzini}}},
  \bibnamefont{et~al.}, \bibinfo{journal}{Astrophysical Journal}
  \textbf{\bibinfo{volume}{710}}, \bibinfo{pages}{133} (\bibinfo{year}{2010}),
  \eprint{\url{http://arxiv.org/abs/0912.3618}}.

\end{thebibliography}

\end{document}